\documentclass[letterpaper,conference]{IEEEtran}
\usepackage[english]{babel}
\IEEEoverridecommandlockouts

\usepackage{microtype}
\usepackage{setspace}
\usepackage{blindtext}
\usepackage{siunitx}
\usepackage[utf8]{inputenc}
\usepackage[T1]{fontenc}
\usepackage{titlecaps}
\Addlcwords{or the with if a of for and to not -off off in by via}
\usepackage{comment}

\usepackage[caption=false]{subfig}
\usepackage{graphicx}
\usepackage{fancyhdr} 
\usepackage{color}
\usepackage{epsfig}
\graphicspath{{Images/}}
\usepackage{tikz}
\usetikzlibrary{patterns,fit,matrix}
\usepackage{tikzscale}
\usepackage{scalerel}
\usetikzlibrary{arrows}
\usetikzlibrary{plotmarks}
\usetikzlibrary{svg.path}
\usetikzlibrary{shapes.multipart}
\usepackage{pgfplots}
\usepgfplotslibrary{fillbetween}
\pgfplotsset{compat=newest}
\pgfplotsset{every axis/.append style={
		label style={font=\Large},
		tick label style={font=\large}  
}}
\tikzstyle{int}=[draw, fill=black!10, minimum size=5em,thick]
\tikzstyle{init} = [pin edge={to-,thick,black}]
\usepackage{booktabs}

\usepackage{array}

\usepackage{enumitem}

\usepackage{amsmath,amssymb,amsfonts,amsthm}
\usepackage{relsize}
\usepackage{nicefrac}
\usepackage{bbm}
\usepackage{mathtools}
\usepackage[mathscr]{eucal}
\usepackage[absolute]{textpos}

\usepackage{url}
\usepackage{hyperref}
\usepackage[capitalize]{cleveref}

\newcommand{\myParagraph}[1]{{\bf #1.}}

\newcommand{\eg}{\emph{e.g.,}\xspace}
\newcommand{\ie}{\emph{i.e.,}\xspace}
\newcommand{\vs}{\emph{vs.}\xspace}

\newcommand{\tradeoff}{latency-accuracy trade-off\xspace}
\newcommand{\homProcPol}{homogeneous sensing policy\xspace}

\newcommand{\Real}[1]{ { {\mathbb R}^{#1} } }

\newcommand{\inv}{^{-1}}

\DeclareMathOperator*{\argmin}{arg\,min}

\newcommand{\tr}[1]{\mathrm{trace}\left(#1\right)}

\newcommand{\var}[1]{\mathrm{Var}\left(#1\right)}

\newcommand{\gauss}{\mathcal{N}}

\newcommand{\x}[1]{x_{#1}}
\newcommand{\w}[1]{w_{#1}}

\newcommand{\yi}[2]{y_{#1}^{(#2)}}
\newcommand{\vi}[2]{v_{#1}^{(#2)}}
\newcommand{\V}[1]{V_{#1}}
\newcommand{\delayRaw}[1][\@empty]{\tau_%
	\ifx\@empty#1 {\text{raw}} \else {{#1,\text{raw}}} \fi}
\newcommand{\delayProc}[1][\@empty]{\tau_%
	\ifx\@empty#1 {\text{proc}} \else {{#1,\text{proc}}} \fi}
\newcommand{\delayComm}[1][]{\delta_{#1}}
\newcommand{\delayCommRaw}[1][\@empty]{\delta_%
	\ifx\@empty#1 {\text{raw}} \else {{#1,\text{raw}}} \fi}
\newcommand{\delayCommProc}[1][\@empty]{\delta_%
	\ifx\@empty#1 {\text{proc}} \else {{#1,\text{proc}}} \fi}
\newcommand{\delayRec}[1]{\Delta_{#1}}
\newcommand{\delayRecRaw}[1][\@empty]{\Delta_%
	\ifx\@empty#1 {\text{raw}} \else {{#1,\text{raw}}} \fi}
\newcommand{\delayRecProc}[1][\@empty]{\Delta_%
	\ifx\@empty#1 {\text{proc}} \else {{#1,\text{proc}}} \fi}

\newcommand{\varRaw}[1][\@empty]{V_%
	\ifx\@empty#1 {\text{raw}} \else {{#1,\text{raw}}} \fi}
\newcommand{\varRawS}[1][\@empty]{v_%
	\ifx\@empty#1 {\text{raw}} \else {{#1,\text{raw}}} \fi}
\newcommand{\varProc}[1][\@empty]{V_%
	\ifx\@empty#1 {\text{proc}} \else {{#1,\text{proc}}} \fi}
\newcommand{\varProcS}[1][\@empty]{v_%
	\ifx\@empty#1 {\text{proc}} \else {{#1,\text{proc}}} \fi}
\newcommand{\samplingSequence}[1]{\mathcal{K}_{#1}}
\newcommand{\measurements}[2][]{\mathcal{Y}_{#2}^{#1}}
\newcommand{\allmeasurements}[1]{\mathcal{Y}_{#1}}

\newcommand{\kell}[2][]{k^{(\ell#1)}_{#2}}
\newcommand{\kL}[1][]{k^{(L)}_{#1}}
\newcommand{\policy}[1][]{\pi_{#1}}
\newcommand{\policyHom}{\pi_\text{hom}}
\newcommand{\decisionHom}[1]{\pi_{\text{hom},#1}}

\newcommand{\sensSet}{\mathcal{V}}

\newcommand{\xhat}[2]{\hat{x}_{#1}^{#2}}
\newcommand{\xtilde}[2]{\tilde{x}_{#1}^{#2}}
\newcommand{\Pmat}[2]{P_{#1}^{#2}}
\newcommand{\pred}[2]{\mathcal{P}^{#2}\left(#1\right)}
\newcommand{\update}[2]{\mathcal{U}\left(#1,#2\right)}

\newcommand{\kalman}[1]{f_{\text{Kalman}}\left(#1\right)}

\newcommand{\epsmax}{\epsilon_0}
\newcommand{\epsmin}{\epsilon_{\text{min}}}
\newcommand{\discFact}{\gamma}
\newcommand{\learnRate}{\alpha}

\theoremstyle{plain}

\theoremstyle{definition}
\newtheorem{definition}{Definition}
\newtheorem{prob}{Problem}
\newtheorem{ass}{Assumption}
\theoremstyle{remark}
\newtheorem{rem}{Remark}

\addto\captionsenglish{\renewcommand{\figurename}{Fig.}}
\addto\captionsenglish{\renewcommand{\tablename}{TABLE}}

\newcommand{\blue}[1]{{\color{blue}#1}}

\newcommand{\linkToPdf}[1]{\href{#1}{\blue{(pdf)}}}
\newcommand{\linkToPpt}[1]{\href{#1}{\blue{(ppt)}}}
\newcommand{\linkToCode}[1]{\href{#1}{\blue{(code)}}}
\newcommand{\linkToWeb}[1]{\href{#1}{\blue{(web)}}}
\newcommand{\linkToVideo}[1]{\href{#1}{\blue{(video)}}}
\newcommand{\linkToMedia}[1]{\href{#1}{\blue{(media)}}}
\newcommand{\award}[1]{\xspace} 
\addto\extrasenglish{}
\addto\extrasenglish{}
\addto\extrasenglish{}

\title{\LARGE\bfseries\titlecap{
A reinforcement learning approach to sensing design in resource-constrained wireless networked control systems}}

\author{{Luca Ballotta$ ^* $, Giovanni Peserico$ ^{*\dagger} $, Francesco Zanini$ ^* $}%
	\thanks{This work has been partially funded by 
		the Italian Ministry of Education, University and Research (MIUR) through the PRIN project no. 2017NS9FEY entitled ``Realtime Control of 5G Wireless Networks: Taming the Complexity of Future Transmission and Computation Challenges'' and through the initiative "Departments of Excellence" (Law 232/2016). The views and opinions expressed in this work are those of the authors and do not necessarily reflect those of the funding institutions.}%
	\thanks{$ ^* $Department of Information Engineering, University of Padova, 35131 Padova, Italy 
		{\tt\small \{luca.ballotta.1, giovanni.peserico, francesco.zanini.3\}@phd.unipd.it}}
	\thanks{$ ^\dagger $Autec s.r.l., 36030 Caldogno, Italy}
	\thanks{Authors are listed in alphabetical order and contributed equally.}
}

\begin{document}
	
	\bstctlcite{MyBSTcontrol}
	\begin{textblock}{20}(-2,0.05)
		\footnotesize
		\centering
		\setstretch{1}
		This paper has been accepted for the 61th IEEE Conference on Decision and Control in Cancun, December 6-9 2022.\\
		Please cite the paper as: L. Ballotta, G. Peserico, F. Zanini,\\
		“\titlecap{a reinforcement learning approach to sensing design in resource-constrained wireless networked control systems}”,\\
		IEEE Conference on Decision and Control, 2022.
	\end{textblock}
	\begin{textblock}{10}(3,15)
		\footnotesize
		\centering
		\setstretch{1}
		\textcopyright 2022 IEEE.  
		Personal use of this material is permitted.  
		Permission from IEEE must be obtained for all other uses, in any current or future media, including reprinting/republishing this material for advertising or promotional purposes, creating new collective works, for resale or redistribution to servers or lists, or reuse of any copyrighted component of this work in other works.
	\end{textblock}
	\maketitle

\begin{abstract}
	In this paper,
	we consider a wireless network of smart sensors (agents) that monitor a dynamical process
	and send measurements to a base station that performs global monitoring and decision-making. 
	Smart sensors are equipped with both sensing and computation,
	and can either send raw measurements or process them prior to transmission.
	Constrained agent resources raise a fundamental \emph{\tradeoff}.
	On the one hand, raw measurements are inaccurate but fast to produce.
	On the other hand, 
	data processing on resource-constrained platforms generates \emph{accurate} measurements at the cost of non-negligible \emph{computation latency}.
	Further, if processed data are also compressed,
	{latency} caused by wireless communication might be higher for raw measurements. 
	Hence, it is challenging to decide when and where sensors in the network should transmit raw measurements or leverage time-consuming local processing.
	To tackle this design problem, we propose a Reinforcement Learning approach
	to learn an efficient policy that dynamically decides when measurements
	are to be processed at each sensor.
	Effectiveness of our proposed approach is validated through a numerical simulation
	with case study on smart sensing motivated by the Internet of Drones.
\end{abstract}

\section{Introduction}

Networked Control Systems have represented a major breakthrough
in control theory and applications,
whereby sensing, computation, and actuation shift from being classically centralized
to decentralized and distributed paradigms.
This empowers multiple agents to coordinate and collaborate towards complex global tasks,
such as management of electricity and energy harvesting in smart grids~\cite{8636257,ERDEM201898},
efficient resource utilization in the Internet of Things and smart agriculture~\cite{9316211,s20072081},
modularization and productivity enhancement in Industry 4.0~\cite{BUENO2020106774,IVANOV2018134,KRUGH201889},
and traffic management enabled by interconnected vehicles~\cite{doi:10.2514/6.2020-0602,9072289}.

Further pushed by recent technological advances on both fronts of embedded electronics,
such as micro controllers and GPU processors~\cite{9613590,9156208},
and powerful communication for massive local networks,
such as 5G~\cite{li20185g,9634111},
the current trend is to rely heavily on 
smart sensors and lightweight devices to share the computational burden across all network resources.
Leading paradigms in this respect, such as edge~\cite{shi2016edge,9156208} and fog computing~\cite{yi2015survey}, and federated learning,
are being investigated across several application domains~\cite{8727940,OGRADY201942}.

However,
emerging edge technologies are still limited compared to powerful servers and cloud resources.
Indeed, resource-constrained devices need to account for several factors
including hardware speed and energy consumption,
whereby local data processing comes at the cost of non-negligible runtime.
In particular, we consider the general scenario where a network of agents,
such as robots or smart sensors,
is coordinated by a central station that receives environmental data to perform remote monitoring or decision-making
on a dynamical process of interest.
Limited resources induce a \emph{\tradeoff} at individual agents (also attention \emph{vs.} precision in robotics):
sensors can either send raw, inaccurate measurements to the base station,
or refine them locally before transmission,
incurring extra \emph{processing delay} due to hardware-constrained computation.
Common options are averaging or filtering of noisy samples,
compression of images or other high-dimensional data~\cite{8280159,9561090},
or descent direction computation in online learning~\cite{9336327,8486403}.
The dynamical nature of the monitored system
makes such delayed processed measurements obsolete, 
so that {sensing design} for multiple, possibly heterogeneous, agents 
becomes nontrivial at network level,
and may require online adaptation of local processing.
In particular, as agents cooperate towards the global task,
the choice of which of them should transmit raw or processed data,
and when should they do that,
induces a challenging design problem.
Also, bandwidth constraints given by wireless channel may
further increase complexity of the design by introducing non-negligible \emph{communication latency}.
Moreover, local computation might also compress collected information,
so that raw measurements might take longer to be communicated to the base station.

\myParagraph{Related literature} 
Control theory traditionally focuses on wireless communication-aware estimation and control,
tailoring unreliability, latency, and packet drop~\cite{4118476,6669629}.
Sensors are usually fixed,
so that the \tradeoff is not addressed,
and local computation need not relate to communication latency.
Similar considerations also apply to literature in sensor selection and resource allocation~\cite{Skelton,9152984,s21227492,s20082187},
which considers sensing design in the presence of budget constraints,
with little attention to impact of delays (or even dynamics) on performance.
Also in recent work on co-design of sensing, communication, and control~\cite{9099596,9654960,9589966},
there is still no unifying framework that ties adaptive local processing and sensing design
with variable computation and communication delays.
Within robotic literature,
computation offloading in cloud robotics targets resource-constrained robots
and wireless communication,
but processing is usually designed for a single robot and may also neglects system dynamics~\cite{chinchali2021network}.
Finally, a body of literature tailored to edge and fog computing studies resource-constrained edge devices
that perform local computation,
but still with little understanding of the actual system dynamics on performance
and main focus on analysis or limited budget~\cite{8757960,8406844}.
In contrast, we address jointly sensing (agent local processing), computation and communication latency, and system dynamics,
within a sensing design framework tailored to optimal performance.

{In~\cite{9137405},
the authors proposed a general model for a processing network,
including impact of computation-dependent delays on system dynamics, 
and provided a heuristic sensing design. 
However, a major limitation of that work is that
the designed sensing policy is static,
\ie agents cannot adapt their processing online,
which may hinder performance, in general. 
Also, it was assumed that sensors could store new incoming samples in an (infinite) buffer
while processing one measurement.
We overcome both restrictions through a novel framework that leverages the core insights presented in~\cite{9137405}.}

\myParagraph{We propose two main contributions}
First, we propose a new model for a \emph{processing network},
which tailors more realistic sampling by resource-constrained agents
that can adapt their local processing online to exploit the \tradeoff.
We address sensing design to manage processing across the network
as an optimization problem that explicitly quantifies impact of computation and latency
on performance (\autoref{sec:setup}). 
To solve this problem, we propose a Reinforcement Learning (RL) approach,
which is detailed in~\autoref{sec:RL-formulation}.
RL, and data-driven methods in general,
are becoming more and more popular in design of networked control systems and edge computing,
because of the challenges of real-world applications~\cite{SHI2020115733,OGRADY201942,8486403}.
Finally, in~\autoref{sec:simulation} we validate our approach with a numerical simulation motivated by the Internet of Drones.

\section{\titlecap{System overview and problem formulation}}\label{sec:setup}

In this section,
we first present the components of a processing network (\autoref{sec:system-model}),
and state the general problem of optimal estimation (\autoref{sec:problem-formulation}).
Then, we propose a tractable version of that problem,
and present a simplified setup with homogeneous sensors (\autoref{sec:problem-relaxation}).

\subsection{System Model}\label{sec:system-model}
We consider a networked control system
where smart sensors 
sample a time-varying signal
and communicate acquired measurements to a common base station.
In the following, we describe in detail the components that model
such a \emph{processing network}.

\myParagraph{Dynamical System}
The process of interest is described by a time-varying discrete-time linear dynamical system,
\begin{equation}\label{eq:stateEquation}
	\x{k+1} = A_k\x{k} + \w{k},
\end{equation}
where $x_k\in\Real{n}$ is the to-be-estimated state of the system,
$A_k\in\Real{n\times n}$ is the state matrix,
and $\w{k}\sim\gauss(0,W_k)$ is Gaussian noise that captures uncertainty in the model.
In view of sensor sampling and subsequent data transmission,
we assume a discretized dynamics with time step $T$
and subscript 
$k\in\mathbb{N}$ meaning the $ k $th discrete time instant $ kT $.
Without loss of \linebreak generality, we set the first instant $ k_0 = 1 $.
The sampling time $T$ represents a suitable time scale for the global monitoring
and, possibly, decision-making task at hand. 

\myParagraph{Smart Sensors}
Process modeled by~\eqref{eq:stateEquation} is measured by $N$ smart sensors (also, sensors or agents)
in the set $ \sensSet = \{1,\dots,N\} $,
which output a noisy sample of the state, 
\begin{equation}\label{eq:sensorMeasurement}
	\yi{k}{i} = \x{k} + \vi{k}{i}, \qquad \vi{k}{i}\sim\gauss(0,\V{i,k})
\end{equation}
where $\yi{k}{i}$ is the measurement collected by the $ i $th sensor at time $ k $, $ i\in\sensSet $,
and $\vi{k}{i}$ is measurement noise.

Agents can either communicate raw measurements
or process collected information locally before transmission.
Decentralized processing is largely employed in modern networked systems,
whereby edge- and fog-computing paradigms
alleviate the computational burden of resource-constrained edge servers or workstations.
Processing may account for
data compression, filtering,
or more complex tasks such as feature extraction in robotic perception or
gradient computation in online learning.
Because of their limited hardware resources,
the above choice generates a \textit{latency-accuracy trade-off} at each agent:
raw measurements are less accurate,
but computation of processed data comes at the cost of extra delay.
In particular, dynamics~\eqref{eq:stateEquation} progressively makes outdated measurements less informative
with respect to the current state of the system,
so that high accuracy alone might not pay off in a real-time monitoring task. 
The ensemble of all choices for local processing at agents (\textit{sensing configurations})
affects global performance in a complex manner,
whereby the presence of multiple agents renders unclear which
of them should rely on local processing
and which ones would be better off transmitting raw measurements.
Indeed,~\cite{9137405} shows analytically that, with identical sensors,
the optimal steady-state configuration is nontrivial.
Also, the 
optimal sensing configuration is time varying, in general. 
We formally model the trade-off at each sensor through the following assumptions.

\begin{ass}[Sensing modes]\label{ass:delays-variances}
	The $ i $th sensor can operate in either \textit{raw} or \textit{processing} mode.
	\emph{Raw measurements} are generated after delay $ \delayRaw[i] $ with noise covariance $\V{i,k}\equiv\varRaw[i]$. 
	\emph{Processed measurements} are generated after \emph{processing delay} $\delayProc[i] $ with noise covariance $\V{i,k}\equiv\varProc[i]$. 
\end{ass}

\begin{ass}[Nominal sampling frequency]\label{ass:sensor-frequency}
	If the $ i $th sensor acquires a sample at time $ k $,
	the next sample occurs at time $ k^+=s_i(k) $,
	\begin{equation}\label{eq:sensor-frequency}
		s_i(k) \doteq \begin{cases}
			k + \delayRaw[i]  & \text{if }\yi{k}{i} \text{ is raw}\\
			k + \delayProc[i] & \text{if }\yi{k}{i} \text{ is processed},
		\end{cases}
	\end{equation}
	\begin{equation}\label{eq:sampling-sequence}
		\begin{gathered}
		    \samplingSequence{i} = \left\lbrace s_i^t\left(k_0\right) \right\rbrace_{t=0}^{\infty} \\
		    \text{s.t. } s_i^{t+1}\left(k\right) = s_i\left(s_i^t\left(k\right)\right); \ s_i^0\left(k_0\right) \doteq k_0,
		\end{gathered}
	\end{equation}
	where 
	$ \samplingSequence{i} $ is the sequence of all sampling instants from $ k_0 $,
	and we denote the $ l $th element of the sequence by $ \samplingSequence{i}[l] $.
\end{ass}

\begin{ass}[Latency-accuracy trade-off]\label{ass:tradeoff}
	For each sensor $ i\in\sensSet $, it holds $ \delayProc[i] > \delayRaw[i] $ and $\varRaw[i]\succ\varProc[i]$.%
	\footnote{Covariance matrices are ordered according to L\"{o}wner order of positive semidefinite matrices.
		Even though this is a partial order, we require it in our model to express the latency-accuracy trade-off unambiguously.}
\end{ass}

In words,~\cref{ass:sensor-frequency} states that sensors have no buffer,
but a new sample is collected only after the previous one has been transmitted (and possibly processed). 
Similarly to~\cite{9137405},
\cref{ass:tradeoff} models high accuracy
through \textquotedblleft small" intensity (covariance) of measurement noise,
\eg uncertainty of $ 1\si{\meter^2} $ of raw \vs $ 0.1\si{\meter^2} $ of processed distance measurements.

\begin{rem}[Multiple processing modes]
	Sensors may have multiple options to process data.
	For example, robots equipped with smart cameras might run different geometric inference algorithms
	exhibiting \tradeoff (in general, anytime behavior~\cite{Zilberstein_1996}).
	While we stick to a single processing mode for the sake of simplicity,
	our model and learning framework in~\autoref{sec:RL-formulation} can
	also be extended in that respect. 
\end{rem}

\myParagraph{Wireless communication}
All sensors transmit over a shared wireless channel.
This introduces \emph{communication latency}
to transport data from sensors to the base station,
due to constraints of wireless transmission.%
\footnote{We assume that communications delays are comparable with $ T $. This is not restrictive as the opposite case is easily modeled by neglecting them.}
Two scenarios are possible: either delays do not depend on agent local processing
(\eg ARMA filtering),
or processed data are compressed (\eg 3D mesh compression).
The latter case may bear a second performance trade-off:
raw data are fast to compute but their transmission takes long,
and vice versa for processed data. 

\begin{ass}[Communication model]\label{ass:comm-delays}
	Packet drop and channel erasure probability are negligible.
	\begin{description}[leftmargin=*]
		\item[No compression.] Raw and processed data are transmitted with communication delay  $ \delayComm[i] \ge 1 $, $ i\in\sensSet $.
		\item[Compression.] Raw measurements are transmitted with communication delay $ \delayComm[i] = \delayCommRaw[i] $ and
		processed measurements with communication delay $ \delayComm[i] = \delayCommProc[i] \le \delayCommRaw[i] $.
	\end{description}
	The \textit{delay at reception} $ \delayRec{i} $ is the overall delay of measurements
	transmitted by the $ i $th sensor to the base station. 
	In particular, $ \delayRecRaw[i] \doteq \delayRaw[i]+\delayCommRaw[i]  $ and $ \delayRecProc[i]\doteq\delayCommProc[i] + \delayProc[i] $.
\end{ass}
\autoref{fig:delayed-meas} illustrates sensing operations with computation delays 
and transmissions with communication delays. 

\myParagraph{Base Station}
Data are transmitted 
to a base station in charge of estimating the state of the system $ \x{k} $ in real time.
This enables global monitoring and possibly decision-making,
\eg remote target tracking or centralized online learning. 
\begin{definition}[Available sensory data]\label{def:data-collected}
	In view of Assumptions~\ref{ass:delays-variances},~\ref{ass:sensor-frequency} and~\ref{ass:comm-delays},
	all sensory data available at the base station at time $ k $ are
	\begin{equation}\label{eq:sequence-measurements}
		\begin{gathered}
			\measurements{k} \doteq {\bigcup_{i\in\sensSet}\bigcup_{l\in\mathbb{N}}}\left\lbrace \left(\yi{\samplingSequence{i}\left[ l \right]}{i},\V{i,\samplingSequence{i}\left[ l \right]}\right) : \samplingSequence{i}\left[ l \right] + \delayRec{i,\samplingSequence{i}\left[l\right]} \le k\right\rbrace, \\
			\delayRec{i,\samplingSequence{i}\left[l\right]} \doteq \begin{cases}
				\delayRecRaw[i]  & \text{if }\yi{\samplingSequence{i}\left[ l \right]}{i} \text{ is raw}\\
				\delayRecProc[i] & \text{if }\yi{\samplingSequence{i}\left[ l \right]}{i} \text{ is processed},
			\end{cases}
		\end{gathered}
	\end{equation}
	where the $ l $th measurement from the $ i $th sensor $ \yi{\samplingSequence{i}\left[ l \right]}{i} $
	is sampled at time $ \samplingSequence{i}\left[ l \right] $ and received after overall delay $ \delayRec{i,\samplingSequence{i}\left[l\right]} $.
\end{definition}
According to~\cref{def:data-collected},
a measurement $ \yi{h}{i} $ is available for 
real-time estimation at time $ k $ if it is successfully delivered to the base station before or at time $ k $,
which entails that it has delay $ \delayRec{i,h} $ through computation and communication.
In~\autoref{fig:delayed-meas}, estimate of $ \x{k_2} $ is computed using
only the first (leftmost) measurement,
which is received 
after delay $ \delayRecProc[i] $.

\subsection{Problem Statement}\label{sec:problem-formulation}

\begin{figure}
	\centering
	\includegraphics[width=\linewidth]{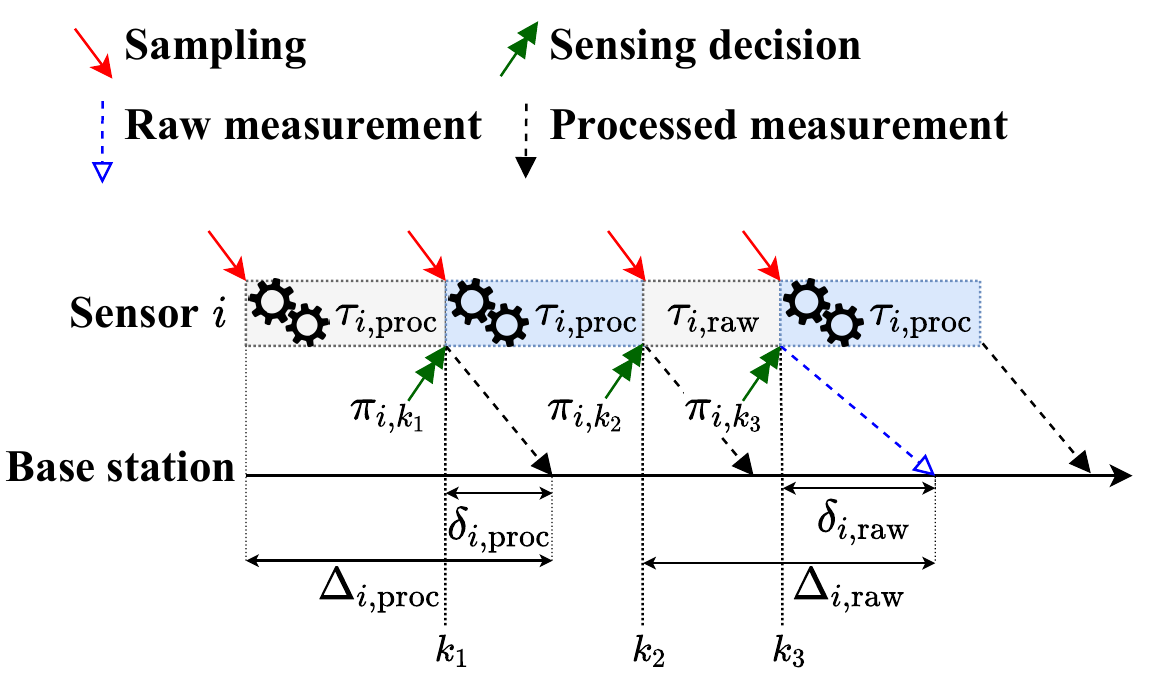}
	\caption{Data are received at the base station after delay induced by computation
	(rectangular blocks) and communication (dashed arrows).
	Local computation at the $ i $th sensor is ruled by sensing policy $ \policy[i] $.
	Given sensing decision $ \policy[i,k_1] = 1 $,
	the sample collected at time $ k_1 $ is processed (processing delay $ \delayProc[i] $),
	transmitted at time $ k_2 = k_1 + \delayProc[i] $
	(communication delay $ \delayCommProc[i] $),
	and received at the base station at time $ k_2 + \delayCommProc[i] = k_1 + \delayRecProc[i] $.
	}
	\label{fig:delayed-meas}
\end{figure}
The trade-offs introduced in the previous section
make the sensing design challenging when this scales to network level. 
In particular, it is not clear which sensors should 
transmit raw measurements,
with poor accuracy and possibly long communication,
and which sensors should refine data locally,
with extra processing delay,
and further, when should they choose either mode.
Henceforth, it is desired to design 
\textit{sensing policies} that decide
on potential processing on-board each sensor in the network.
We denote the policy for the $ i $th sensor by $ \policy[i] \doteq \{ \policy[i,k]\}_{k\in\mathbb{N}} $.
Each \textit{sensing decision} $ \policy[i,k] $ is a binary variable:
if $ \policy[i,k] = 0 $, the measurement collected by sensor $ i $ at time $ k $ is transmitted raw,
while if $ \policy[i,k] = 1 $ the measurement is processed.
Note that the sampling sequence $ \samplingSequence{i} $ itself will depend on policy $ \policy[i] $,
according to evolution~\eqref{eq:sensor-frequency},
cf.~\autoref{fig:delayed-meas}.

The estimate of $ \x{k} $, denoted by $ \xhat{k}{} $,
is computed via Kalman predictor,
which is the optimal estimator for linear systems driven by Gaussian noise.
It can be shown, \eg via standard state augmentation,
that the Kalman predictor is optimal even with delayed or dropped measurements,
whereby it suffices to ignore updates with measurement
associated with missing data.
Out-of-sequence arrivals
can be handled by recomputing the most recent update steps starting from the latest arrived measurements,
or by more sophisticated techniques~\cite{kalmanDelays,kalmanDelays1}.
Let $\xtilde{k}{}\doteq\x{k}-\xhat{k}{}$ denote the estimation error associated with $\xhat{k}{}$,
and let $ \Pmat{k}{}\doteq\var{\xtilde{k}{}} $ the error covariance matrix.
We state the sensing design as an optimal estimation problem.

\begin{prob}[Sensing Design for Processing Network]\label{prob:optimal-estimation}
	Given system~\eqref{eq:stateEquation}--\eqref{eq:sensorMeasurement} with Assumptions~\ref{ass:delays-variances}--\ref{ass:comm-delays},
	find the optimal sensing policies $ \{\policy[i]\}_{i\in\sensSet} $,
	so as to minimize the time-averaged estimation error variance with horizon $K$,
	\begin{equation}\label{eq:prob-optimal-estimation}
		\begin{aligned}
			\argmin_{\policy[i]\in\Pi_i,i\in\sensSet}& &&\frac{1}{K}\sum_{k=k_0}^K \tr{\Pmat{k}{\policy}}\\
			\text{s.t.}& &&\Pmat{k}{\policy} = \kalman{\measurements[\policy]{k}}\\
			& &&\Pmat{k_0}{\policy} = P_0,
		\end{aligned}
	\end{equation}
	where the Kalman predictor $ \kalman{\cdot} $ computes at time $ k $ state estimate $ \xhat{k}{\policy} $ and error covariance matrix $ \Pmat{k}{\policy} $
	using data $ \measurements[\policy]{k} $ available at the base station according to $ \policy \doteq \{\policy[i]\}_{i\in\sensSet} $,
	and $ \Pi_i $ gathers all causal sensing policies of the $ i $th sensor.
\end{prob}

\subsection{\titlecap{sensing policy: a centralized implementation}}\label{sec:problem-relaxation}

\cref{prob:optimal-estimation} is combinatorial and 
does not scale with the size of the system. 
This raises a computational challenge in finding efficient sensing policies,
as the search space may easily explode. 
On the one hand,
the possible simultaneous sensing configurations do not scale with the number of sensors,
\eg $ 10 $ sensors yield $ 2^{10} = 1024 $ possible sensing configurations at each sampling instant.
On the other hand,
\cref{prob:optimal-estimation} also requires to design the policy for each sensor,
which is a combinatorial problem by itself that scales exponentially with the time horizon $ K $.
Furthermore,
each sensing policy $ \policy[i] $ not only affects data delay and accuracy,
but also determines the sampling sequence $ \samplingSequence{i} $ for the $ i $th sensor (cf.~\eqref{eq:sensor-frequency}),
augmenting the search space to all possible sampling instants.

To make the problem more tractable,
we restrict to a simple setup
that greatly reduces computational complexity
while still enabling useful insights and relevance to applications. 
We then propose a simplification of the original~\cref{prob:optimal-estimation}
to be tackled by our learning method
(\autoref{sec:RL-formulation}). 

\begin{rem}[Complexity with multiple processing modes]
	In the general case 
	where the $ i $th sensor has $C_i$ processing modes, 
	there are $\prod_{i\in\sensSet}C_i$ sensing configurations in total.
\end{rem}

\subsubsection{Homogeneous Sensors}\label{sec:hom-sens}

In this case, 
all sensors have identical sensing modes and noise statistics,
\begin{equation}\label{eq:hom-sensors}
    \yi{k}{i} = \x{k} + \vi{k}{i}, \qquad \vi{k}{i}\sim\gauss(0,\V{k}),
\end{equation}
with
$ \V{k} = \varRaw $ or $ \V{k} = \varProc $ for raw and processed data, respectively.
Further, all sensors have computational delays $ \delayRaw $ and $ \delayProc $,
and communication delays $ \delayCommRaw $ and $ \delayCommProc $
($ \delayComm $ in case of no compression),
for raw and processed data, respectively.
Such \textit{homogeneous sensor} scenario
models the special but relevant case where sensors are interchangeable.
This happens, \eg with sensors monitoring
temperature in large plants or
with smart cameras extracting high-level information
such as object pose.

\subsubsection{Problem Simplification}
In this scenario,
one just needs to decide \textit{how many}, 
rather than \emph{which}, 
sensors shall process at each time.
Hence, 
the amount of sensing configurations drops to $ N+1 $,
greatly reducing the search space.\footnote{
	In the general case with $ C $ processing modes for each sensor, the total amount of configurations becomes $\binom{N+C-1}{C-1}$.}
Accordingly,
we shift attention to
a centralized \emph{\homProcPol}.

\begin{definition}[Homogeneous sensing policy] \label{def:hom_case}
	A \emph{\homProcPol} is a sequence $ \policyHom = \{\decisionHom{\ell}\}_{\ell=1}^L $,
	where each \emph{homogeneous decision} $ \decisionHom{\ell} \in \{0,\dots,N\} $ is taken at time $ \kell[]{} $.
	If $ \decisionHom{\ell} = p $, then $ p $ sensors process their measurements
	and $ N-p $ send them raw between times $ \kell[]{} $ and $ \kell[+1]{} $.
	Without loss of generality, we set $ k^{(1)} = k_0 $ and $ \kL < K $.
\end{definition}

\begin{figure}
	\centering
	\includegraphics[width=\linewidth]{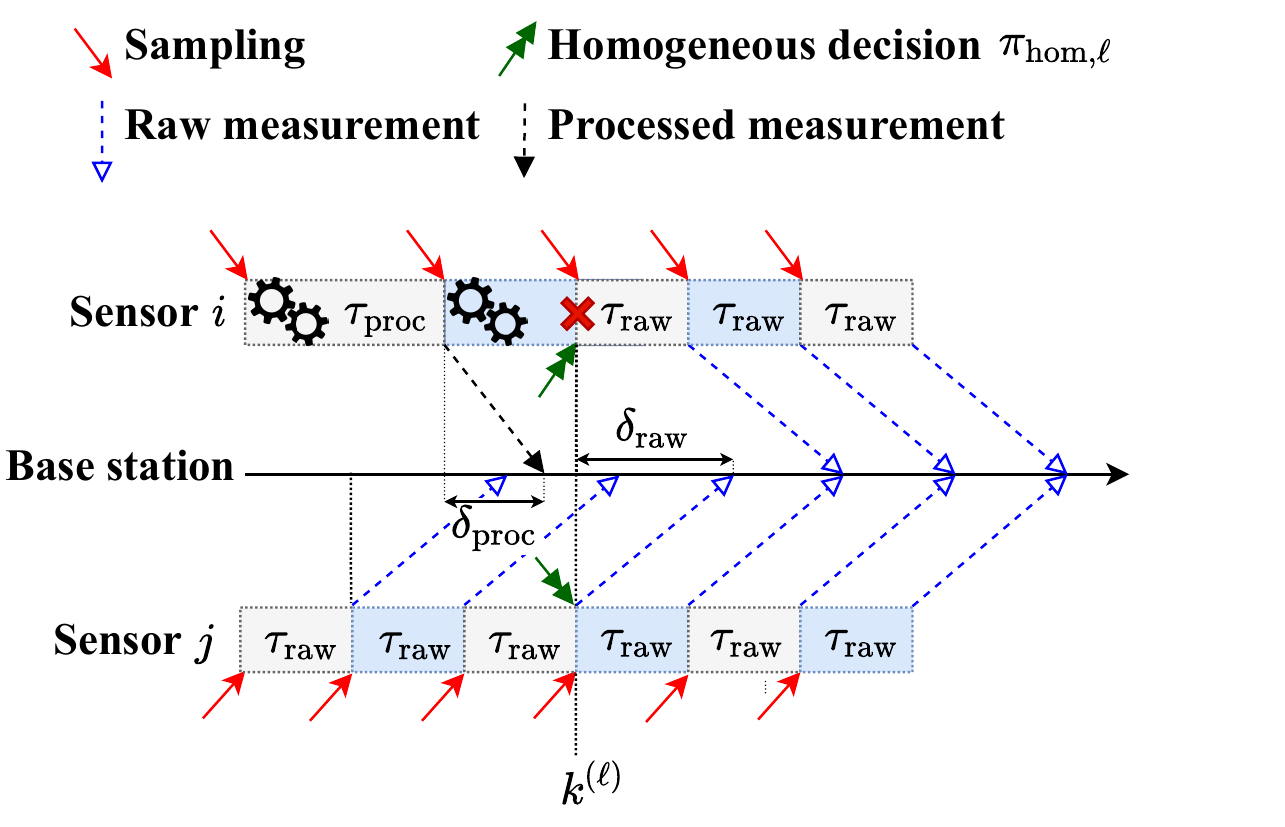}
	\caption{Homogeneous sensors ruled by policy $ \policyHom $.
		As decision $ \decisionHom{\ell} $ is communicated, 
		the $ i $th sensor disregards current measurement (red cross) and
		immediately switches to raw mode, acquiring a new sample.
	}
	\label{fig:delayed-meas-centr}
\end{figure}

In words,
the base station decides for all sensors at predefined time instants,
which is convenient both in practical applications
and to set up our learning procedure.
However,
frequency of decisions may be arbitrary, 
in general.

Centralized decisions may be communicated while some sensors
are still processing their data
According to standard real-time control~\cite{5510096,s22041638},
we assume what follows.

\begin{ass}[Sampling frequency with \homProcPol]\label{ass:decision-time-centralized-policy}
	Decision $ \decisionHom{\ell} $ affects the minimum amount of sensors possible.
	When a sensor is commanded to switch mode,
	the current measurement is immediately discarded
	and sampling restarts with the new mode.
	Hence, given a measurement sampled at time $ k $,
	dynamics~\eqref{eq:sensor-frequency} becomes
	\begin{equation}\label{eq:sensor-frequency-centralized-policy}
		s_i^h(k) \doteq \begin{cases}
			\min\left\lbrace k + \delayRaw,\kell{}\right\rbrace  & \text{if }\yi{k}{i} \text{ raw}\\
			\min\left\lbrace k + \delayProc,\kell{}\right\rbrace & \text{if }\yi{k}{i} \text{ processed},\\
		\end{cases}
	\end{equation}
	where $ \yi{k}{i} $ is sampled before decision $ \decisionHom{\ell} $ is communicated
	and it is discarded if it is not transmitted by time $ \kell{} $.
	Further, we require $ \kell[+1]{} \ge \kell{} + \delayProc, \ell = 1,\dots,L-1 $.
\end{ass}

In view of~\cref{ass:decision-time-centralized-policy},
the available sensory data at the base station
now exclude
measurements discarded by overlapping decisions.
Formally, 
for the $ i $th sensor, 
such data
are raw (resp., 
processed) measurements $ \yi{k}{i} $ sampled under decision $ \decisionHom{\ell-1} $
such that $ k + \delayRaw > \kell{} $ (resp., 
$ k + \delayProc > \kell{} $),
\ie their generation ends after a \textit{different} mode is imposed at time $ \kell{} $ by decision $ \decisionHom{\ell} $.
Figure~\ref{fig:delayed-meas-centr} illustrates the new sampling dynamics, 
with the second measurement of the $ i $th sensor discarded at time $ \kell{} $.
To avoid excessive notational burden, 
we simply write $ \measurements[\policyHom]{k} $
to denote available data at the base station at time $ k $ according to~\eqref{eq:sensor-frequency-centralized-policy} and
the discussed data dropping mechanism imposed by policy $ \policyHom $,
that prunes some measurements from the set $ \measurements[]{k} $.

In this scenario, 
\cref{prob:optimal-estimation} simplifies as follows.

\begin{prob}[Sensing Design for Homogeneous Processing Network]\label{prob:optimal-estimation-hom}
	Given system~\eqref{eq:stateEquation},~\eqref{eq:hom-sensors} with Assumptions~\ref{ass:delays-variances}--\ref{ass:decision-time-centralized-policy}
	and sequence $ \{\kell{}\}_{\ell=1}^L $, $ \kell[+1]{} \ge \kell{} + \delayProc \,\forall\ell<L$, $ L < K $,
	find the optimal \homProcPol $ \policyHom $,
	with $ \ell $th decision $ \decisionHom{\ell} $ scheduled at time $ \kell{} $,
	so as to minimize the time-averaged estimation error variance with horizon $ K $,
	\begin{equation}\label{eq:prob-optimal-estimation-hom}
		\begin{aligned}
			\argmin_{\policyHom\in\Pi_\text{hom}}& &&\frac{1}{K}\sum_{k=k_0}^K \tr{\Pmat{k}{\policyHom}}\\
			\text{s.t.}& &&\Pmat{k}{\policyHom} = \kalman{\measurements[\policyHom]{k}}\\
			& &&\Pmat{k_0}{\policyHom} = P_0,
		\end{aligned}
	\end{equation}
	where the Kalman predictor $ \kalman{\cdot} $ computes at time $ k $ state estimate $ \xhat{k}{\policyHom} $ and error covariance matrix $ \Pmat{k}{\policyHom} $
	using data $ \measurements[\policyHom]{k} $ available at the base station according to $ \policyHom $,
	and $ \Pi_\text{hom} $ gathers all causal homogeneous sensing policies.
\end{prob}

\begin{rem}[Heterogeneous sensors]\label{rem:het-sens}
	\cref{prob:optimal-estimation-hom} can be easily generalized to heterogeneous sensors,
	whereby decisions are centralized and the base station schedules \emph{which} sensors process data
	at each decision $ \policy[\ell] $.
	However, 
	while such implementation eliminates complexity given by asynchronous mode switching of agents,
	the action space may explode with the amount of sensors.
	To keep computational burden reasonable,
	here we focus on the homogeneous case.
	Future work will focus on compute-efficient strategies for heterogeneous networks,
	possibly via multi-agent Reinforcement Learning.
\end{rem}

\newcommand{\lr}{\left(}
\newcommand{\rr}{\right)}
\newcommand{\ls}{\left[}
\newcommand{\rs}{\right]}
\newcommand{\lb}{\left\lbrace}
\newcommand{\rb}{\right\rbrace}

\section{Reinforcement Learning Framework}\label{sec:RL-formulation}

By assuming complete knowledge of delays 
and measurement noise covariances affecting sensors in the different modes,
both~\cref{prob:optimal-estimation} and~\ref{prob:optimal-estimation-hom} become analytically tractable.
However, the computation of the exact minimizer requires to keep track of all starts and stops of data transmission for each sensor,
resulting in a cumbersome procedure which admits no closed-form expression,
and requires to solve a combinatorial problem which scales poorly with the amount of sensors. 

Moreover, in real-life scenarios, the assumptions considered in the formulation of the problem may be too conservative,
and the latter method does not allow for relaxations. 
Indeed, as long as either delays or covariances are not explicitly known or have some variability,
\ie they can be modeled by proper random variables,
the minimization becomes intractable.
This is true even if the expectations of these random variables are known,
since the dynamics in~\cref{prob:optimal-estimation-hom} leads nonetheless to a non-linear behavior for the quantity of interest.

As a matter of fact, it is clear from~\cref{prob:optimal-estimation-hom} that the covariance matrix $P_k$ is strongly affected by the sensing configuration,
\ie which sensors process their data and which transmit them raw.
The problem of choosing the optimal sensing policy in order to minimize the uncertainty of estimation is tackled through a Reinforcement Learning algorithm, 
because it allows much greater flexibility in the formulation and assumptions of the problem.

\subsection{General Scenario}\label{sec:RL-general-scenario}

The Reinforcement Learning framework~\cite{Sutton1998} consists of an agent interacting
with an environment without having any prior knowledge of how the latter works and how its actions may impact on it.
By collecting results of the interactions, which are expressed through a reward signal,
the agent will learn to maximize this user-defined quantity, under the action of the unknown environment.
This can be formalized with the help of the Markov Decision Process framework,
in which the tuple $\left< \mathcal{S}, \mathcal{A}, \mathcal{P}, r, \gamma \right>$ characterize all needed elements,
\ie the state space $\mathcal{S}$;
the set of all possible actions $\mathcal{A}$;
the transition probabilities $\mathcal{P}$ regulating the underlying system, which may also be deterministic;
the reward function $r$, which is assumed to be a function of the state and the action;
and the discount factor $\gamma$, weighting future rewards.
The agent selects actions to be performed on the environment through a \textit{control policy},
which is here assumed to be a deterministic function $ \pi $ mapping states to actions, $\pi: \mathcal{S} \rightarrow \mathcal{A}$.
The goal of the learning procedure would be to identify an optimal policy $\pi^*(\cdot)$, 
\ie a map allowing to achieve the highest possible return $G$ over the long run,
according to the unknown transitions imposed by the environment.
The return is simply defined as the expectation of the sum of rewards in an episode of $K$ steps, 
\ie $G = \mathbb{E}\left[ \sum_{k=1}^K r_k \right]$. 

The general problem can then be written as
\begin{equation} \label{eq:RL-problem}
	\begin{aligned}
		\max_{\policy \in \Pi}& &&\mathbb{E}\ls\sum_{k=0}^{K-1}\gamma^k r(s_k, a_k)\rs \\
		\text{s.t.}& 			&&s_{k+1}^{\policy} = g\lr s_k, a_k \rr = g\lr s_k, \pi\lr s_k\rr\rr \doteq g_\pi\lr s_k\rr,
	\end{aligned}
\end{equation}
for a generic state-transition function $g: \mathcal{S}\times\mathcal{A} \rightarrow \mathcal{S}$ 
representing environment dynamics that embeds the transition probabilities $\mathcal{P}$,
where $ \Pi $ is the space of all admissible policies.
One of the simplest and most popular approaches to solve the maximization in~\eqref{eq:RL-problem} is the Q-learning algorithm \cite{Watkins1992}.

Q-learning is a \textit{model-free} algorithm which finds an optimal policy to maximize the expected value of the return, 
without learning an explicit model of the environment. 
This is done by focusing solely on the \textit{values} of actions at different states, which correspond to the return. 

In the tabular setting this method actually builds a lookup table with the action-value for each state-action pair,
and updates it with an \textit{optimistic} variant of the temporal-difference error~\cite{10.1023/A:1022633531479} at every step, 
weighted by a learning rate $\alpha$.

The action-value function is defined as
\begin{equation}
	Q^\pi(s, a) = \mathbb{E}_\pi\ls\sum_{k=0}^{K-1}\gamma^k r(s_k, a_k) \Bigg\vert s_k = s, a_k=a\rs,
\end{equation}
and it expresses the expected return obtained by performing action $a$ at state $ s $ and then following policy $\pi$ afterwards. \\
The Q-learning algorithm iteratively updates an estimate of the action-value function $\hat{Q}$, replacing the current estimate with the one given by the best performing action and bootstrapping for the future predictions. \\
By defining the optimistic variant of the temporal-difference error as
\begin{equation}
	\zeta^\pi_k \lr s, a\rr = r\lr s, a\rr + \gamma\max_{a'}\ls \hat{Q}_k \lr s',a'\rr\rs - \hat{Q}_k \lr s,a \rr,
\end{equation}
the update for the Q-learning algorithm is given by
\begin{equation}
	\hat{Q}_{k+1}\lr s,a\rr = 
	\begin{cases}
		\hat{Q}_k\lr s,a\rr + \alpha \, \zeta^\pi_k \lr s, a\rr, & k<K\!-\!1 \\
		\lr1-\alpha\rr \hat{Q}_k\lr s,a\rr + \alpha \, r\lr s_k, a_k\rr, & \text{o.w.}
	\end{cases}
\end{equation}
The Q-function is updated at each step $k = 1, 2, \dots$ of one episode. 
The policy deployed in the environment to run the episodes and collect rewards 
is chosen as the $\epsilon$-greedy policy according to the current lookup table, 
emphasizing the explorative behavior of the algorithm.

In a finite MDP, this approach is known to converge to the \emph{optimal} action-value function under the standard Monro-Robbins conditions \cite{NIPS1993_5807a685}.

\subsection{Optimizing Latency-Accuracy Trade-off}\label{sec:RL-tradeoff}

With regard to~\cref{prob:optimal-estimation}, 
policy $\policy[i]$ is made of categorical variables that take as many values as possible sensor modes,
and it completely characterizes the potential for intervention in the measuring process of the $ i $th sensor. 
The simplification due to the homogeneous scenario
allows to consider 
a single policy $ \policyHom : \mathcal{S} \rightarrow \mathcal{A} $
described by a sequence of integers (see Definition~\ref{def:hom_case})
which indicate how many sensors are required to process their measurements at each time $ \kell{} $. 
This means that an action $ a $ corresponds to the amount of sensors
that are required to process their measurements after $ a $ is applied,
and it lies in the set $ \mathcal{A} = \{0,\dots, N\} $.

Since in the latter case the aim is to minimize the estimation error covariance through
the choice of the sensing configuration overtime (see~\cref{prob:optimal-estimation-hom}), 
a straightforward metric to be chosen as reward function is the negative trace of matrix $\Pmat{k}{\policyHom}$, 
which evolves according to the Kalman predictor with delayed updates (Appendix~\ref{app:kalman-filter}). 
The base station is allowed to change sensing configuration (corresponding to a new action) at each time $\kell[]{}$,
therefore a natural way of defining the reward is to take the negative average of the trace of the covariance
during the interval between times $\kell[]{}$ and $\kell[+1]{}$,
so that the base station can appreciate the performance of a particular sensing configuration in that interval. 

This leads to the following specialization of the maximization~\eqref{eq:RL-problem},
\begin{equation}
	\begin{aligned}
		\max_{\policyHom \in \Pi_{\text{hom}}}& && \mathbb{E}\ls\sum_{\ell=1}^{L}\gamma^\ell \sum_{k=\kell{}}^{\kell[+1]{}}\frac{\tr{\Pmat{k}{\policyHom}}}{\kell[+1]{} - \kell{}}\rs \\
		\text{s.t.}& &&\Pmat{k}{\policyHom} =\kalman{\measurements[\policyHom]{k}},
	\end{aligned}
\end{equation}
with $ k^{(L+1)} \doteq K $.
The quantity of interest is the trace of the error covariance and thus a straightforward approach would suggest to take $\mathcal{S} = \mathbb{R}^+$, 
however to keep the Q-learning in a tabular setting the state space has been discretized through the function $\mathfrak{d}: \mathbb{R}^+\rightarrow\mathbb{N}^+$. 
In particular, $\mathfrak{d}\ls\cdot\rs$ is built by running a full episode for every possible static policy ${\policyHom}(\cdot) \equiv a$ (denoted in what follows by "All-$a$"),
and selecting $M$ bins in such a way that the relative frequency of the collected states result the same for each bin. 
This results in the state space $ \mathcal{S} = \{0,\dots,M\}$,
and allows for a fair representation of the values of $ \Pmat{k}{\policyHom} $ observed along episodes.

The agent is only aware of the bin of $ \Pmat{\kell{}}{} $ at each time $\kell{}$,
and based on that it outputs a sensing configuration $ a\in\mathcal{A} $ through $\policyHom\lr\cdot\rr$,
given by $a_\ell = \policyHom\lr \mathfrak{d}\ls \tr{\Pmat{\kell{}}{}} \rs\rr$.

\section{\titlecap{numerical simulation}}\label{sec:simulation}

We illustrate the effectiveness of our approach
with a numerical experiment involving four agents with homogeneous computation.
As a motivating scenario, we consider four drones that are tracking an object (\eg vehicle)
which is moving on a planar surface.
Akin~\cite{9137405}, we assume that the vehicle moves approximately at constant speed,
and we model it as a stochastically forced double integrator~\cite{ABDESSAMEUD2010812}
with variance of process noise associated with velocity equal to $ 10^{-1} $.
As for sensing, the four drones are equipped with cameras and can either transmit raw images,
that provide a noisy estimate of the vehicle state,
or further process camera frames to detect the vehicle onboard
and transmit more accurate information.
In particular, drones perform object detection with a neural network model amenable for mobile platforms,
that guarantees fair accuracy to comply with real-time inference requirements on resource-constrained platforms.
We assume cameras sampling at 25FPS and choose processing delays based on real-world data~\cite{raspberryPi,s19153371}.
Further, we set $ \varRaw = \varRawS I $ and $ \varProc = \varProcS I $.
As for communication,
we assume that transmitted raw camera frames are small, to save bandwidth and transmission power,
and consider equal communication delay $ \delayComm $ for raw and processed data.
Simulation parameters are summarized in~\cref{table:params-sensing}.
Finally, we set $ \Pmat{0}{} = 10I $ as initial error covariance of the Kalman filter.

The base station makes a new decision every $ 500\si{\milli\second} $,
and the predefined sequence of decision instants is thus $ k^{(1)} = 0 $, $ k^{(2)} = 50 = 0.5\si{\second}\dots $
We call a time interval between two consecutive decisions a \textit{window},
and optimize over a total of ten windows (decisions),
with time horizon $ K = 500 = 5\si{\second} $.
\begin{table}
	\begin{center}
		\caption{System parameters used in simulation.}
		\label{table:params-sensing}
		\begin{tabular}{cccccc}
			\toprule
			$ T $ 					&  $ \delayRaw $ 	  	  & $ \delayProc $			 	&  $ \varRawS $ &  $ \varProcS $ 		& $ \delayComm $\\
			$ 10 $\si{\milli\second}& $ 40 $\si{\milli\second}& $ \delayRaw+100 $\si{\milli\second}	& $ 10 $ 	    & $ 1 $					& $ 10 $\si{\milli\second}\\
			\bottomrule
		\end{tabular}
	\end{center}
\end{table}
\begin{table}
	\begin{center}
		\caption{Hyperparameters used in Q-learning.}
		\label{table:params-RL}
		\begin{tabular}{cccccc}
			\toprule
			$ M $ & $ \learnRate $ 	& $ \epsmax $ & $ \epsmin $ & $ \epsilon_t $ 												   & $ \discFact $\\
			$ 5 $ & $ 0.01 $ 		& $ 0.9 $	  & $ 0.1 $     & $ \max\left\lbrace{\dfrac{\epsmax}{\sqrt{t}}},\epsmin\right\rbrace $ & $ 0.99 $ 	   \\
			\bottomrule
		\end{tabular}
	\end{center}
\end{table}

\cref{table:params-RL} shows the hyperparameters used in the Q-learning.
In particular, we force exploration at the beginning to avoid strong biases in the Q-table during the first iterations,
and gradually decrease the exploration parameter $ \epsilon_t $ at each episode $ t $
(note that one episode corresponds to one full time horizon, \ie $10$ windows).
We let the $ \epsilon $-greedy policy explore constantly also after many episodes (as $ \epsilon_t $ settles to $ \epsmin $)
to make sure that the action space is sufficiently explored for all states,
given that the reward difference between similar actions is very small at certain states.
Also, we set a small learning rate $ \alpha $ to balance the large difference in rewards that the same state-action pair
may experience at difference windows
(recall that the estimated return is given by the sum of rewards till the end of one episode).	

\begin{table}
	\begin{center}
		\caption{Costs and returns with Q-learning and static policies.}
		\label{table:policies-costs}
		\begin{tabular}{c|c|c||c|c|c}
			\toprule
			$ \policyHom $ & Cost & Return & $ \policyHom $ & Cost & Return\\
			\toprule
			Q-l & \boldmath $ 6.373 $ & \boldmath $ -60.997 $    & All-2 & $ 6.403 $ & $-61.291$\\
			All-$ 0 $ & $ 6.442 $ & $-61.685$ & All-3 & $ 6.506 $ & $ -62.290 $\\
			All-$ 1 $ & $ 6.374 $  & $-61.013$ &  All-4 & $ 6.759 $ & $-64.772$ \\
			\bottomrule
		\end{tabular}
	\end{center}
\end{table}

\begin{table}
	\begin{center}
		\caption{Actions applied by Q-learning policy along optimized horizon.}
		\label{table:Q-learning-policy}
		\begin{tabular}{c|c|c|c|c|c|c|c|c|c|c}
			\toprule
			Window & $ 1 $ & $ 2 $ & $ 3 $ & $ 4 $ & $ 5 $ & $ 6 $ & $ 7 $ & $ 8 $ & $ 9 $ & $ 10 $\\
			\hline
			Action & $ 1 $ & $ 0 $ & $ 1 $ & $ 1 $ & $ 1 $ & $ 1 $ & $ 1 $ & $ 1 $ & $ 1 $ & $ 1 $\\
			\bottomrule
		\end{tabular}
	\end{center}
\end{table}

\begin{figure}
	\centering
	\includegraphics[width=.9\linewidth]{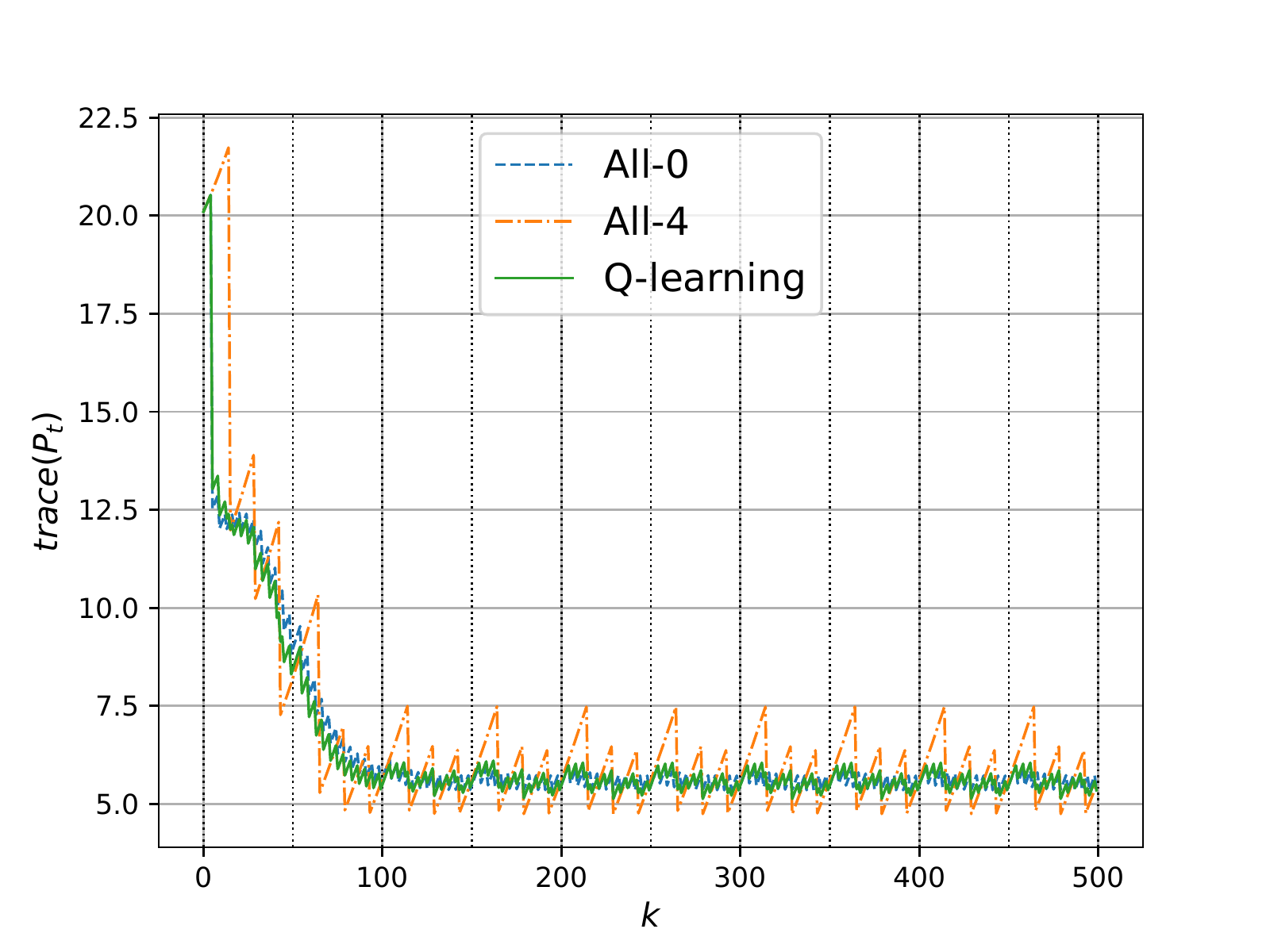}
	\caption{Time evolution of $ \tr{P_k} $: Q-learning \vs common (static) policies.
		Windows are marked by vertical lines at times $ k = 0, 50, 100,\dots $
	}
	\label{fig:compare-policies}
\end{figure}

\begin{figure}
	\centering
	\includegraphics[width=.9\linewidth]{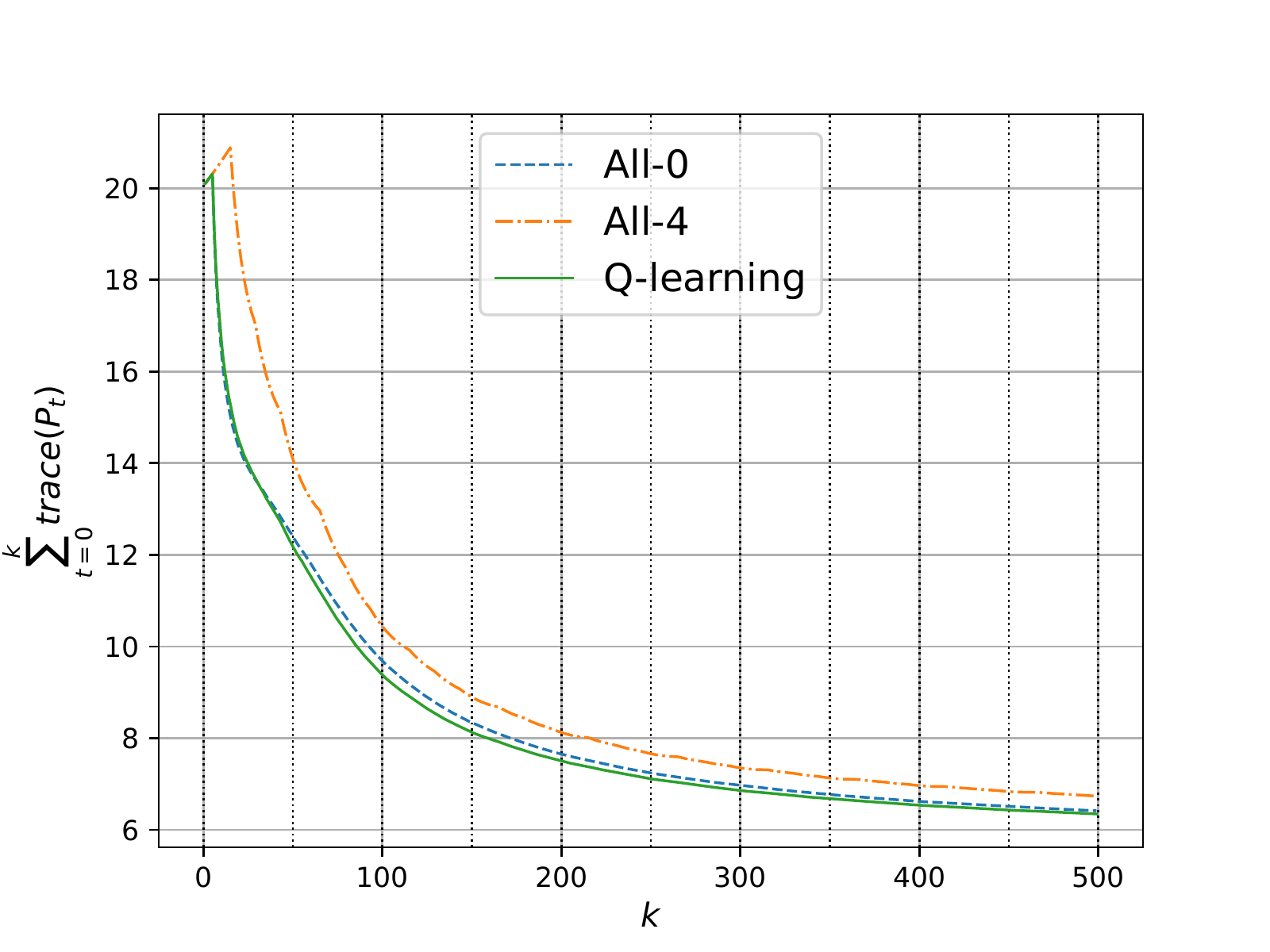}
	\caption{MA of $ \tr{P_k} $: Q-learning \vs common static policies.}
	\label{fig:compare-policies-MA}
\end{figure}

\cref{table:policies-costs} shows the performance obtained with the static policies
and the one learned by Q-learning (Q-l),
where "Cost" refers to the value of the objective function in~\eqref{eq:prob-optimal-estimation-hom}.
Static policy All-$ a $ applies action $ a $ at all windows,
\eg All-$ 2 $ commands $ 2 $ drones to perform object detection.
The learned policy Q-l outperforms all static ones,
resulting in both smaller estimation cost and larger return (bold).
In particular, it always applies action $ 1 $ with the exception of the second window,
when $ \Pmat{k}{} $ is still undergoing the transient phase
and it chooses action $ 0 $, \ie all drones send raw images (\cref{table:Q-learning-policy}).
As the time evolution of $ \tr{\Pmat{k}{}} $ in~\autoref{fig:compare-policies}
and its moving average in~\autoref{fig:compare-policies-MA} show,
this is crucial as more frequent measurements are exploited to drop the variance during the second window.
Then, the best action on the long run $ a = 1 $, \ie making one single drone process, is recovered till the end of the horizon.

\begin{rem}[Nontrivial optimal design]
	All-$ 0 $ and All-$ 4 $ represent two common design choices:
	the former neglects processing, so that drones simply send raw images,
	while the latter fully exploits local processing by all drones at all times.
	Hence, we achieve two relevant insights.
	Firstly, none of such standard choices is optimal,
	as also remarked in~\cite{9137405}.
	In particular, forcing local processing at all agents may be suboptimal in the presence of computational delays,
	as shown by the high cost of All-$ 4 $ in~\cref{table:policies-costs} and~\autoref{fig:compare-policies}.
	Secondly, the optimal policy is in fact time-varying,
	and exploits the \tradeoff in different ways at different times.
\end{rem}

\section{Conclusion}\label{sec:conclusion}

Motivated by resource-constrained networked systems, 
we have proposed a sensing design that addresses impact on performance
of computation and communication latency and of local processing onboard agents.
We have tackled this problem via Q-learning, showing that the learned policy can outperform
common design choices, as well as static policies.
	

\appendix[\titlecap{kalman predictor with delayed updates}]\label{app:kalman-filter}

Assume that all received measurements available at time $ k $ 
are ordered in time as follows,
\begin{equation}\label{eq:all-measurements-sorted}
	\allmeasurements{k} = \left\lbrace \left(y_{k_0},V_{k_0}\right),,\dots,\left(y_{k_M},V_{k_M}\right) \right\rbrace, \quad k_i < k_{i+1},
\end{equation}
where $ M $ is the total amount of measurements and $ k_M < k $.
Note that the following procedure can handle an out-of-sequence measurement sampled at time $ k_0 $ (oldest in~\eqref{eq:all-measurements-sorted})
and received at time $ k $.
For the sake of clarity, in~\eqref{eq:all-measurements-sorted} we have deliberately omitted subscripts and superscripts
associated to sensing mode and sensor.
The estimation error covariance given by Kalman predictor starting from $ P_{k_0} $ is computed by
\begin{equation}\label{eq:Kalman-filter}
	\Pmat{k}{} = \pred{\update{...\update{\pred{\update{P_{k_0}}{V_{k_0}}}{k_0:k_1}}{V_{k_1}}{...}}{V_{k_M}}}{k_M:k}
\end{equation}
where the multi-step open-loop update is
\begin{equation}\label{eq:Kalman-filter-prediction-update}
	\begin{aligned}
		&\pred{\Pmat{k_l}{}}{k_i:k_j} = \mathcal{P}_{k_j}\circ\dots\circ\mathcal{P}_{k_i}(\Pmat{k_l}{}), 
		\ \pred{\Pmat{k_l}{}}{k_i:k_i} \doteq \Pmat{k_l}{}\\
		&\mathcal{P}_{k_i}\left(\Pmat{k_l}{}\right) \doteq A_{k_i}\Pmat{k_l}{}A_{k_i}^\top + W_{k_i},
	\end{aligned}
\end{equation}
with $ k_i \le k_j $, and the update with measurement is
\begin{equation}\label{eq:Kalman-filter-update}
	\update{\Pmat{k_i}{}}{\V{k_i}{}} = \left((\Pmat{k_i}{})\inv + \left(\V{k_i}{}\right)\inv\right)\inv.
\end{equation}
	


\end{document}